\documentclass[aip, amsmath, amssymb, reprint]{revtex4-2}
\usepackage{graphicx}
\usepackage{hyperref}
\usepackage{natbib}
\usepackage{color}

\begin{document}
\title{Intensity correlations in perturbed optical vortices: Diagnosis of the topological charge}
\author{Patnala Vanitha} \email{vanitha\_patnala@srmap.edu.in}
\author{Bhargavi Manupati}
\author{Salla Gangi Reddy} \email{gangireddy.s@srmap.edu.in}
\affiliation{SRM University -- AP, Amaravati, Mangalagiri, Andhra Pradesh, 522502, India.}

\author{Venkateswarlu Annapureddy}
\affiliation{Department of Physics, National Institute of Technology, Tiruchirappalli, Tamil Nadu 620015, India}

\author{Shashi Prabhakar} \email{shaship@prl.res.in}
\author{R. P. Singh} \email{rpsingh@prl.res.in}
\affiliation{Quantum Science and Technology Laboratory, Physical Research Laboratory, Ahmedabad, India 380009.}
\date{\today}

\begin{abstract}
We propose and experimentally verify a novel scheme for diagnosing the order of a perturbed optical vortex using its 2-D spatial autocorrelation function. The order of a vortex was found to be equal to the number of dark rings or zero points present in the correlation function. We have provided a compact analytical expression for the correlation function in the form of Laguerre polynomials. Further, we have utilized the divergence of the first zero point of Laguerre polynomials upon propagation to obtain information about higher-order spatial modes, and compared it with our experimental results. It has been shown that the accuracy of obtained information can further be enhanced by increasing the collection area of scattered vortex beams. These results find applications in free-space optical and satellite communication as the proposed technique is alignment-free, and information can be obtained even with a small portion of the scattered beam.
\end{abstract}

\maketitle

Optical vortices (OV) are screw dislocations in optical fields and have zero intensity at their center. These beams are known for their helical wave-fronts and carry an orbital angular momentum of $\ell\hbar$ per photon, where $\ell$ is the topological charge or order that quantifies the number of phase helices per wavelength \cite{Beijersbergen}. Optical vortices are realized in both coherent and partially coherent fields. The coherent optical vortices can be observed directly, whereas, in partially coherent beams, these vortices can be observed in correlation functions and are also known as coherence vortices \cite{visser, jesus, vanitha}. Both coherent and partial coherent beams have gained a lot of interest in recent years due to their applications such as optical manipulation and communication \cite{Beijersbergen, liu, tabosa, wang}.

A random light granular pattern can be obtained by scattering a coherent light beam through a rough surface, such as a ground glass plate (GGP). The generated random pattern is known as optical speckles, which are formed due to the interference of many scattered waves arising from different inhomogeneities of the rough surface \cite{goodman}. In the past, we have studied the scattering of the optical vortices and observed that the speckle size decreases with the order, which is attributed to the increase in the area of illumination on the rough surface with order \cite{higher, div}. The random optical fields are increasing their interest in science and technology due to their applications in communication, cryptography, and optical manipulation \cite{david, kumar, heeman, fercher}. 

For the applications in free-space optical communication using spatial modes, one needs to propagate these modes for longer distances. After propagating through the channel, the mode information gets disturbed, and one needs to find the mode information of these perturbed beams. Although there are many techniques to find the order of a higher-order coherent optical vortex \cite{prabhakar2011revealing, pravin, courtial, hu, sgreddy}, they are not suitable for partially coherent or incoherent vortices. A limited number of techniques are available for finding the order of a partially coherent vortex beam \cite{cleberson, alves, dong}.

In this work, we propose and implement a novel scheme for finding the order of perturbed optical vortices using a 2-D spatial autocorrelation function and its divergence with propagation distance. We have also provided a detailed mathematical analysis of the 2-D spatial coherence function of scattered optical vortices along with an exact in-line expression for their divergence of zero-points present in it. Further, we validate our theoretical findings with the experimental results. We provide the level of accuracy for measuring the topological charge by considering the deviation between experimental and theoretical results.

We start our theoretical analysis with the field distribution of a Laguerre-Gaussian beam with an azimuthal index $\ell$ and zero radial index, which can be written mathematically as 
\begin{eqnarray}
    E_0(\rho,\theta,0)\propto \rho^{\vert{\ell}\vert}e^{-\frac{\rho^2}{\omega_0^2}}e^{-i\ell\theta},
    \label{eq1}
\end{eqnarray}
where $\omega_0$ is the beam waist and ($\rho, \theta$) are cylindrical coordinates at the vortex generation plane. The scattering of OV beams through a ground glass plate (GGP) can be well modelled with $\delta-$correlated random phase function $\exp(i\Phi)$, where $\Phi$ varies randomly from $0-2\pi$. $\Phi$ can be obtained by taking a 2-D convolution between a random spatial function and a Gaussian correlation function. The field distribution $U(\rho, \theta)$ of speckles obtained after scattering through the GGP is given by 
\begin{eqnarray}
    U(\rho, \theta)=e^{i\Phi(\rho, \theta)}E(\rho,\theta).
    \label{eq2}
\end{eqnarray}
Here, the autocorrelation function for the exponential phase factor is a Dirac-delta function at the source plane $(\rho,\theta)$ and can be represented as
\begin{eqnarray}
    \langle e^{i(\Phi(\rho_1,\theta_1)-\Phi(\rho_2,\theta_2))}\rangle = \delta(\rho_1-\rho_2)\delta(\theta_1-\theta_2),
    \label{eq3}
\end{eqnarray}
where $\langle a \rangle$ represents the ensemble average of parameter $a$. 

The autocorrelation function of scattered light, i.e. speckle patterns, is defined as
\begin{eqnarray}
    \Gamma(r_1,\varphi_1;r_2,\varphi_2)=\langle U_1(r_1\varphi_1)U_2^*(r_2,\varphi_2)\rangle,
    \label{eq4}
\end{eqnarray}
where $r, \varphi$ are coordinates at the detection plane.

The field amplitude at the detection plane can be evaluated with the help of Fresnel’s diffraction integral for the field amplitude at the source plane in cylindrical coordinates \cite{acevedo}.
\begin{eqnarray}
    U(r,\varphi,z)&=&\frac{e^{ikz}}{i\lambda z}\int \rho d\rho \int d\theta U(\rho, \theta) \nonumber \\
   &\times& e^{\left(\frac{ik}{2z}[\rho^{2}+r^{2}-2\rho r\cos(\theta-\varphi)]\right)}.
    \label{eq5}
\end{eqnarray}
Using Eq. \ref{eq5} in Eq. \ref{eq4}, we obtained the mutual coherence function as
\begin{eqnarray}
    \Gamma(r_1,\varphi_1;r_2\varphi_2)&=&\frac{1}{\lambda^2 z^2}e^{\left(\frac{ik}{2z}[r_1^{2}-r_2^{2}]\right)} \int \rho_1 d\rho_1 \int d\theta_1 \nonumber \\ &\times& \int \rho_2 d\rho_2
    \int d\theta_2 \langle U_1(\rho_1,\theta_1) U_2^*(\rho_2,\theta_2)\rangle \nonumber \\
    &\times& e^{\frac{ik}{2z}\left[\rho_1^{2}-\rho_2^{2}-2\rho_1 r_1\cos(\theta_1-\varphi_1) +2\rho_2 r_2\cos(\theta_2-\varphi_2)\right]}
    \label{eq6}
\end{eqnarray}
which is a four-fold integral, that includes the cross-correlation of the field at plane $(\rho,\theta)$, namely, $\langle U_1(\rho_1,\varphi_1) U_2^*(\rho_2,\varphi_2)\rangle$. In addition, using Eq. \ref{eq2} and Eq. \ref{eq3} the cross-correlation at the source plane $(\rho,\theta)$ can be written as
\begin{eqnarray}
    \langle U_1(\rho_1,\theta_1) U_2^*(\rho_2,\theta_2)\rangle   
    &=& E_1(\rho_1,\theta_1) E_2^{*}(\rho_2,\theta_2)\nonumber \\ &&\delta(\rho_1-\rho_2)\delta(\theta_1-\theta_2).
    \label{eq7}
\end{eqnarray}
Ideally, the random phase screens such as GGP can be modelled using the $\delta$-correlated phase function. After using the same in the above equation, we get
\begin{eqnarray}
    \langle U_1(\rho_1,\theta_1) U_2^*(\rho_2,\theta_2)\rangle=E(\rho,\theta) E^{*}(\rho,\theta).
    \label{eq8}
\end{eqnarray}
The four-fold integral mentioned in Eq. \ref{eq6} can be reduced to two fold-integral with the help of Eq. \ref{eq8} and the properties of the Dirac-delta function, which can be written as
\begin{eqnarray}
    \Gamma(r_1,\varphi_1;r_2\varphi_2)&=&\langle U_1(r_1,\varphi_1) U_2^*(r_2,\varphi_2)\rangle \nonumber \\
    &=&\frac{1}{\lambda^2 z^2}e^{\frac{ik}{2z}\left[r_1^{2}-r_2^{2}\right]} \int \rho d\rho \int d\theta E(\rho,\theta) \nonumber \\ & \times& E^{*}(\rho,\theta) e^{\frac{-ik}{z}\rho[r_1\cos(\theta-\varphi_1) -r_2\cos(\theta-\varphi_2)]},
    \label{eq9}
\end{eqnarray}
where
\begin{eqnarray}
    \Delta r \cos(\varphi_s-\theta)&=&\left[\left( r_1 \cos\left( \varphi_1\right) -r_2 \cos\left( \varphi_2\right)\right)\cos\theta\right] \nonumber \\ &+&\left[\left( r_1 \sin\left( \varphi_1\right) -r_2 \sin\left( \varphi_2\right)\right)\sin\theta\right],
    \label{eq10}
\end{eqnarray}
and $\Delta r^2=r_1^2+r_2^2-2r_1r_2 \cos\left(\varphi_2-\varphi_1\right)$. Equation \ref{eq9} can be simplified further as 
\begin{eqnarray}
    \Gamma_{12}\left( \Delta r\right)&=& \frac{1}{\lambda^2 z^2} e^{\left(\frac{ik}{2z}[r_1^{2}-r_2^{2}]\right)}\int\int \vert E_0(\rho,\theta,0)\vert^2 \nonumber \\ & \times& e^{-\frac{ik}{z}\left(\rho \Delta r \cos(\varphi_s-\theta)\right)} \rho d\rho d\theta.
    \label{eq11}
\end{eqnarray}
Using Eq. \ref{eq1}, we get the incident field information as
\begin{eqnarray}
    \vert E_0(\rho,\theta,0)\vert^2=\rho^{2\vert{\ell}\vert} e^{-\frac{2\rho^2}{\omega_0^2}}.
    \label{eq12}
\end{eqnarray}
After inserting the above field information in autocorrelation function and using the Anger-Jacobi identity \cite{ryzhik}, we get that 
\begin{eqnarray}
    \Gamma_{12}\left( \Delta r\right)& = & \frac{\pi \omega_0^{2\vert {\ell}\vert +2} e^{\frac{ik}{2z}\left( r_1^2-r_2^2\right) } }{{2^{\vert {\ell}\vert +1}} \lambda^2 z^2 } \nonumber \\ & \times & e^{\left(-\frac{k^2 \omega_0^2 \Delta r^2}{8z^2}\right)} L_{\vert{\ell}\vert}\left(\frac{k^2 \omega_0^2 \Delta r^2}{8z^2}\right) 
    \label{eq17}
\end{eqnarray}
The autocorrelation function of perturbed optical vortices depends on the order $\ell$ and the propagation distance $z$. The number of zero points in the Laguerre polynomial is equal to order $\ell$, and one can extract the information about the spatial mode. It is known that the Fourier transform of the incident field can represent the autocorrelation function and the Fourier transform of the coherent optical vortex has $\ell$ number of zero points.

However, the experimental observation of zero points in the correlation function of scattered vortices with higher topological charges is difficult due to the reduction in correlation strength with $\Delta r$, the spatial separation distance between two images taken for correlation. To avoid this difficulty, we have considered the first zero of the correlation function and its variation with propagation for diagnosing the topological charge. The position of the first zero-point or ring of the correlation function can be obtained by using the roots of Laguerre polynomials and given by
\begin{equation}
    c_\ell = \frac{k^2\omega_0^2 \Delta r^2}{8z^2},
\end{equation}
where $c_\ell$ is the lowest root of the Laguerre polynomial. The correlation length is defined as the value of $\Delta r$ at which we have zero-correlation and can be determined as
\begin{equation}
    \Delta r_0 = \frac{\sqrt{8c_\ell}}{k\omega_0}z.
\end{equation}
From the above equation, it is clear that the correlation length is directly proportional to the propagation distance $z$. Now, the divergence of this first zero point is defined as the rate of change of its variation with propagation distance \cite{div}. Mathematically it can be written as
\begin{equation}
    \alpha = \frac{\partial \Delta r_0}{\partial z}= \frac{\sqrt{8c_\ell}}{k\omega_0}.
    \label{eq:div}
\end{equation}
Experimentally, one can find the divergence as slope of $\Delta r_0$ versus $z$, and this quantity is independent of propagation distance. We utilize this information with the proper usage of input parameters such as wavelength and beam waist to find the topological charge of a perturbed vortex. 

\begin{figure}[h]
   \includegraphics[width=7.5cm]{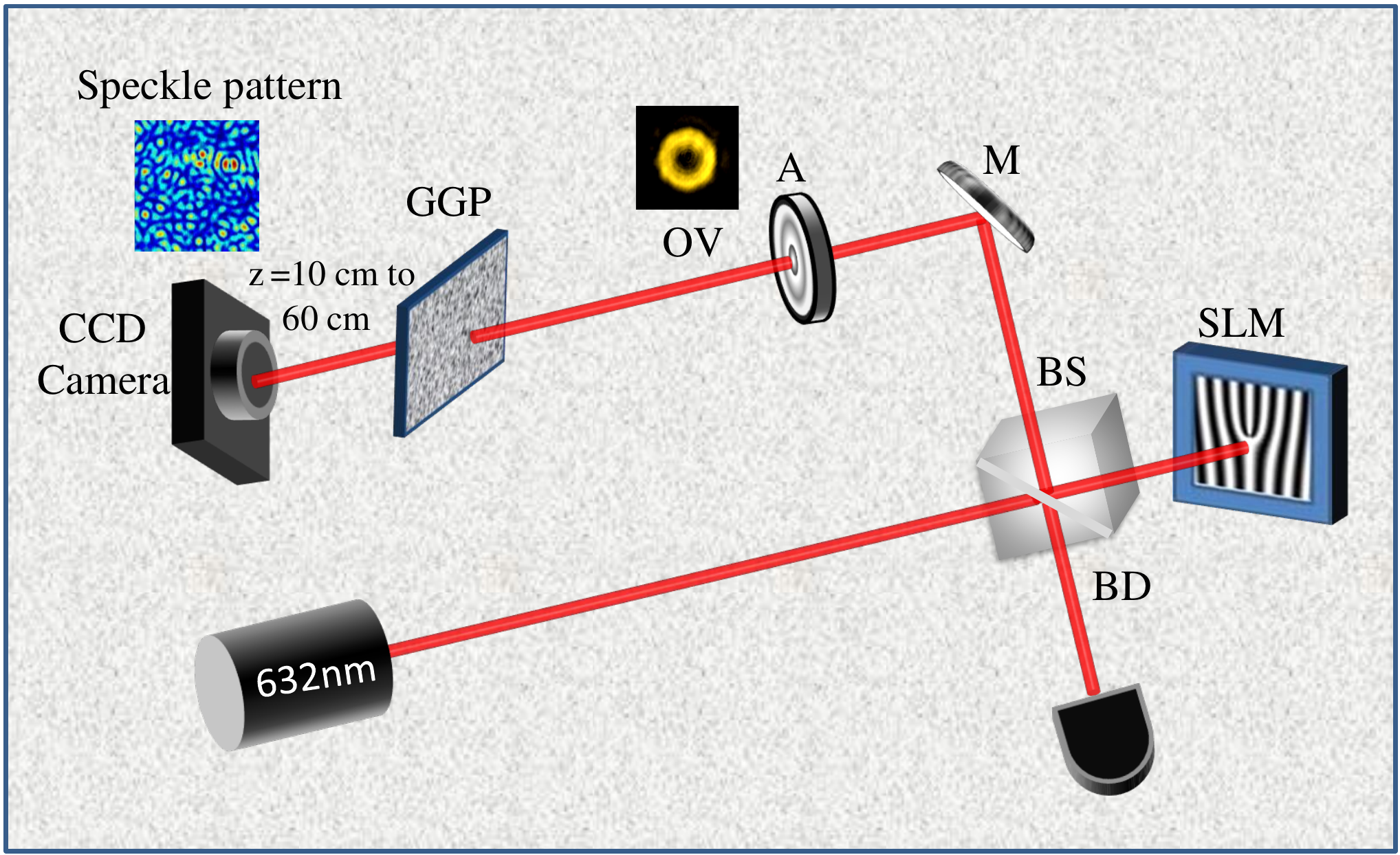}
    \caption{(Colour online) Experimental setup for the generation of optical vortices and the corresponding speckle patterns.}
    \label{fig:expt}
\end{figure}

The experimental setup for generating the optical vortices is shown in Fig. \ref{fig:expt}. A He-Ne laser beam of wavelength 632.8 nm and power $<$5 mW illuminates the spatial light modulator (SLM) containing a computer-generated hologram to generate optical vortices in different diffraction orders. The desired vortex mode is selected from an aperture (A) and scattered through a ground glass plate (Thorlabs DG10-600). The random fields, i.e. speckle patterns obtained after the scattering, have been recorded using a CCD camera with a pixel size of 3.45 $\mu$m at different distances from 10 cm to 30 cm in the steps of 5 cm from the GGP. We have used the experimentally obtained beam waist value of 0.55 mm for all our theoretical findings.

\begin{figure}[h]
   \includegraphics[width=7.5cm]{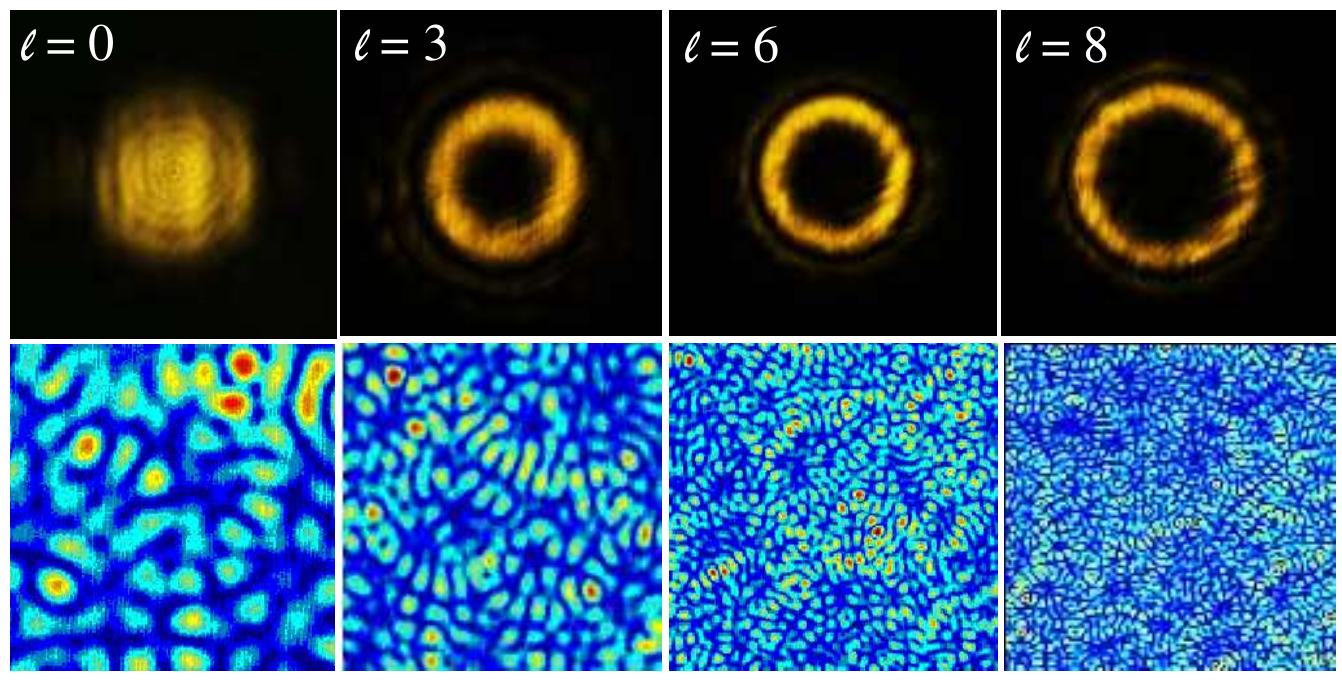}
    \caption{(Colour online) The experimental images for the intensity distributions of Optical vortices (top) and the corresponding speckle patterns (bottom) for orders $\ell=0,3,6,8$ from left to right.}
    \label{fig:optical vortex}
\end{figure}

We start our experiment by recording the intensity distributions of the optical vortex beam at the plane of GGP for all orders from $\ell=0-8$. Then, we scattered these light beams through a GGP and recorded the corresponding speckle patterns, which have been shown in Fig. \ref{fig:optical vortex}. It is evident from the figures that the size of the speckles decreases with the increase in order, which is in agreement with our previously published results \cite{higher}. 

\begin{figure}[h]
   \includegraphics[width=7.5cm]{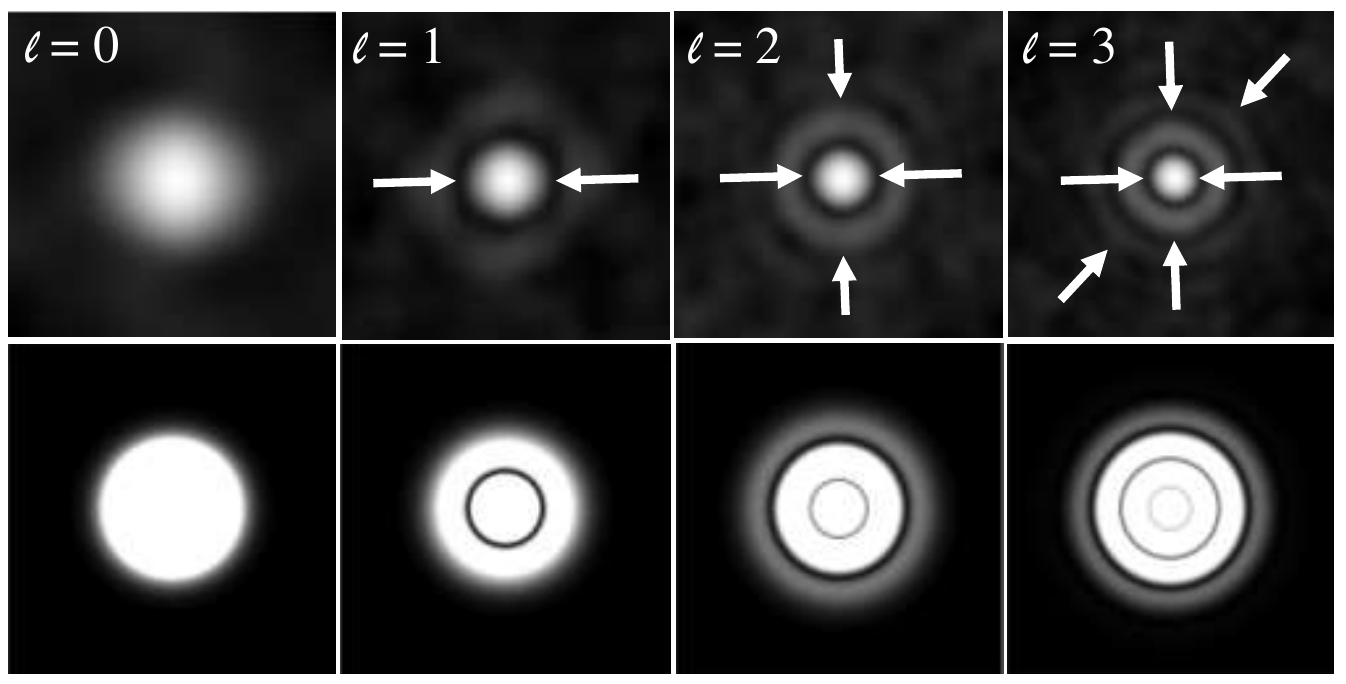}
    \caption{(Colour online) Experimental (top) and theoretical (bottom) generation of spatial correlation field for a LG beam with $p=0$ and $\ell=0,1,2,3$ from left to right. The arrow marks provide a better view for identifying the rings.}
    \label{fig:rings}
\end{figure}

After recording the speckle patterns, we calculated their 2-D spatial autocorrelation function using Matlab, which will be further used to extract the information about the spatial mode. Figure \ref{fig:rings}, shows the experimental (top) and theoretical (bottom) images for the autocorrelation function of speckle patterns for the optical vortex beams of the order $\ell=$0-3. We diagnose the order or topological charge of a given spatial mode by the number of dark rings present in the autocorrelation function. For better observation of the rings, we have inserted the arrow marks at the position of the rings. This method works very well for vortices with low topological charges; however, for higher-order vortices, the observation of rings is difficult due to the decreased correlation strength with spatial separation distance ($\Delta r$). The observation of number of rings in the auto-correlation function may find applications in free-space optical communication using vector vortex beams where we consider the superposition of two orthogonal spatial modes of low topological charge with mutually independent polarization. This technique is alignment-free as the autocorrelation is independent of alignment.

\begin{figure}[h]
    \includegraphics[width=7.5cm]{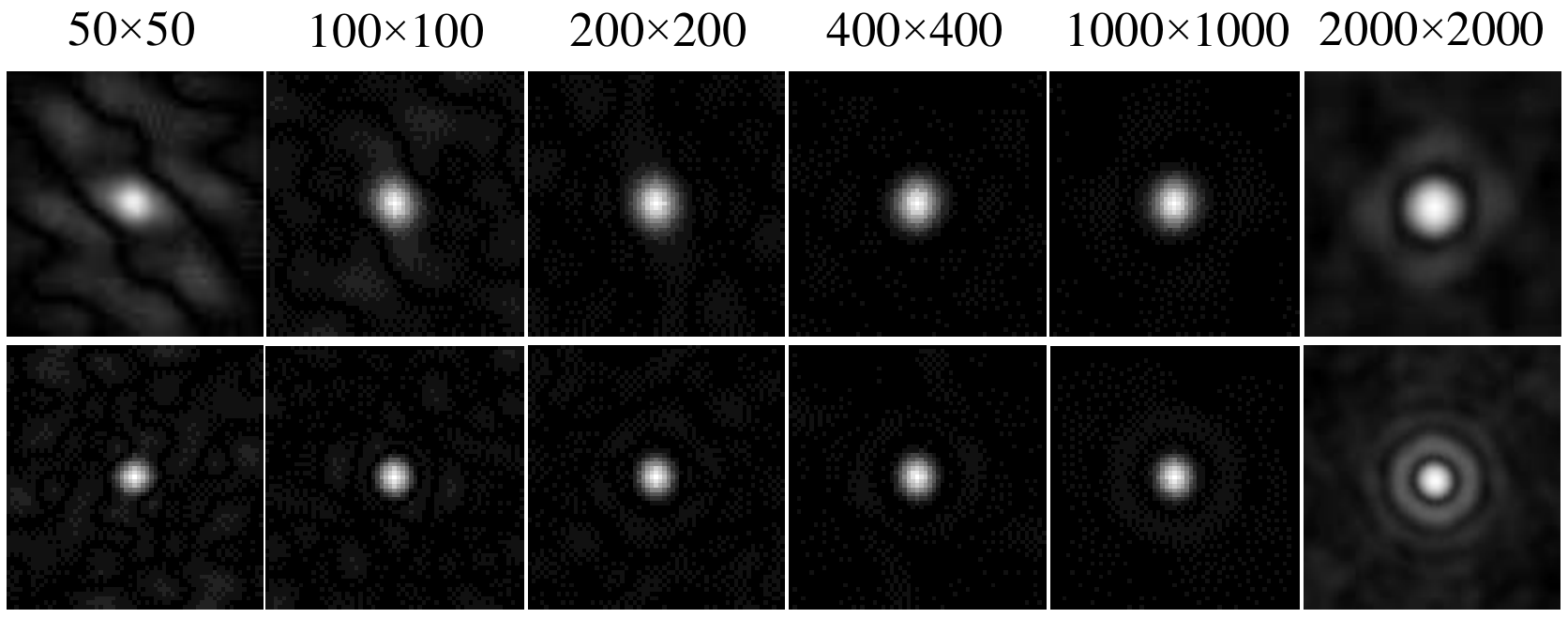}
    \caption{(Colour online) 2-D spatial autocorrelation function of scattered optical vortex beams of order $\ell=1$ (top) and $\ell=3$ (bottom) with a different number of pixels as written at the top.}
    \label{fig:diffpixelrings}
\end{figure}

Further, we examined the size of the speckle pattern collected through a CCD camera on the autocorrelation function and the results for $\ell=$ 1 and 3 have shown in Fig. \ref{fig:diffpixelrings}. The observation of an exact number of rings depends on the total number of pixels, and for the first order, we can observe the ring when we take a minimum of 200$\times$200 image. With the increase in the number of pixels used to find the 2-D autocorrelation function, one can enhance visibility and accuracy in finding the mode information.

\begin{figure}[h]
    \includegraphics[width=7.5cm]{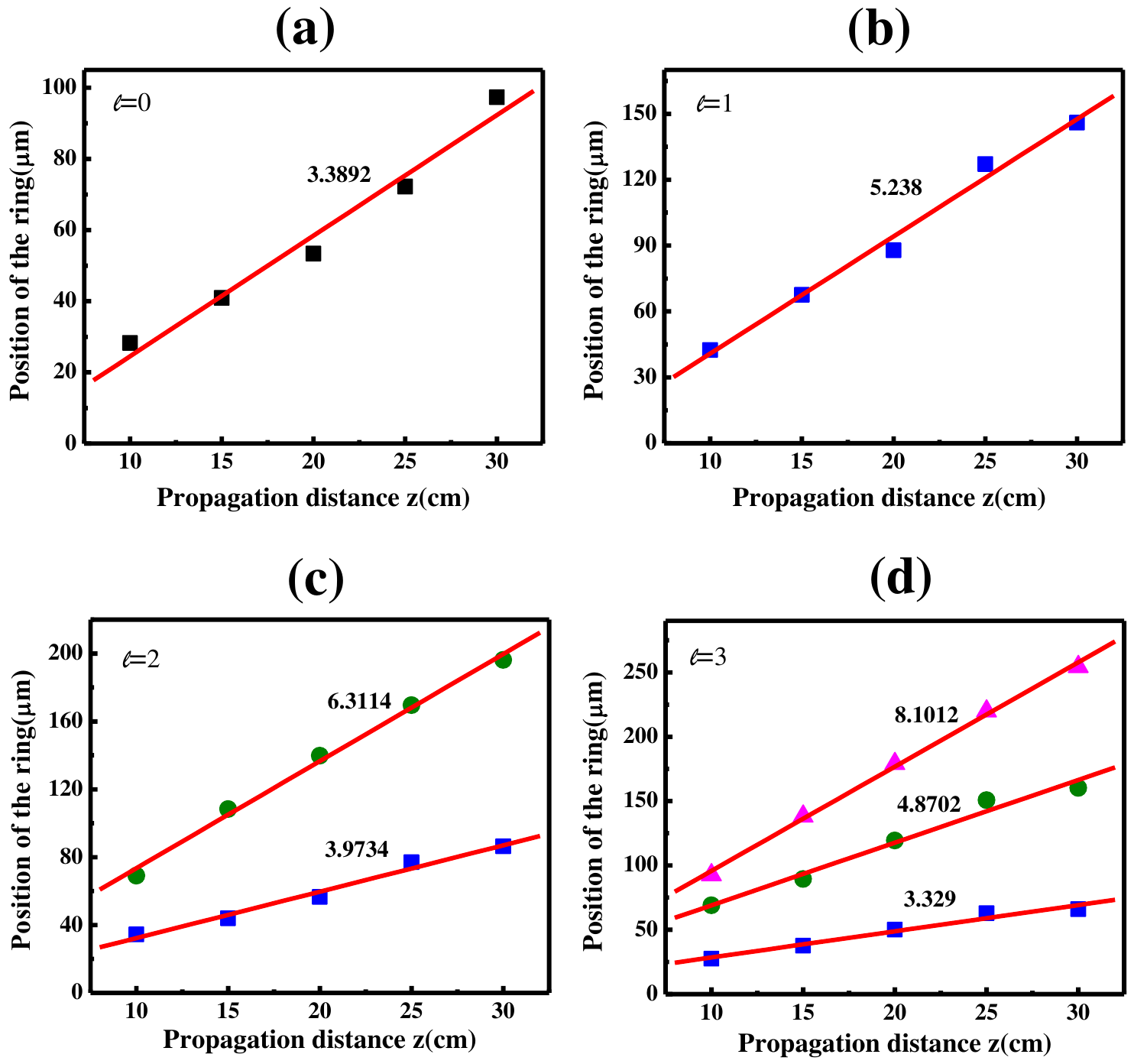}
    \caption{(Colour online) The variation of the width of Gaussian function and the position of zero-points with propagation distance for different orders from $\ell$=0-3 along with their divergence values given at the top of data points.}
    \label{fig:ringsposition}
\end{figure}

We can collect a maximum of 2500$\times$2500 image through a CCD camera, and observing the dark rings for higher orders may still be challenging. Further, we have utilized the variation of the position of dark rings with propagation distance $z$. Figure \ref{fig:ringsposition} shows the variation of the position of dark rings with propagation distance for different orders $\ell=$0-3. For $\ell=$0, we have considered the width of the Gaussian function (with respect to $I/e^2$) as beam diameter, and for all other orders, we have considered intensity zeros as the position of dark rings or zero points. For $\ell=$1, we get one ring, for $\ell=$2 we get two rings and for $\ell=$3, we get three rings, and their positions have been determined at all propagation distances. We observed that the zero point varies linearly with the propagation distance, and we have defined their divergence as the rate of change of zero-point position with propagation distance, i.e. the slope of a line drawn for zero-point position $\Delta r_0$ versus propagation distance $z$. For the outer ring, the divergence increases with order and is shown on top of the plots, whereas the divergence for lower rings decreases with the order, as shown in figure \ref{fig:ringsposition}. 

\begin{figure}[h]
    \includegraphics[width=7.5cm]{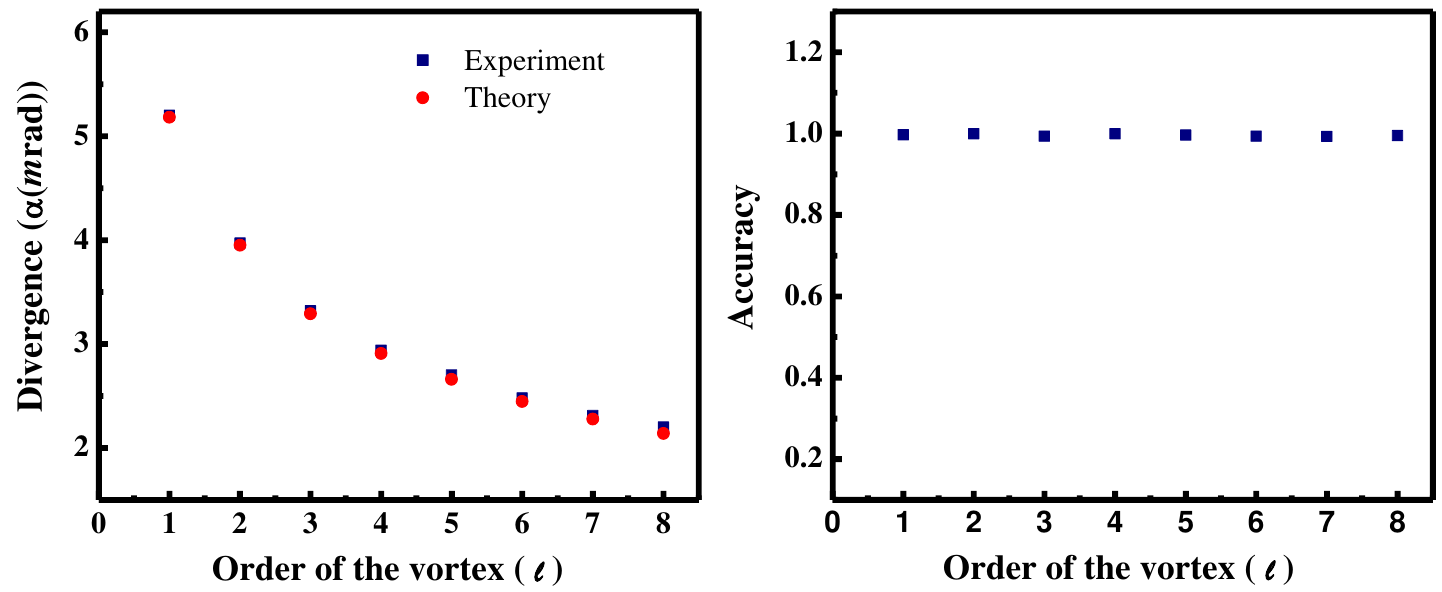}
    \caption{(Colour online) (a) Experimental (blue) and theoretical (green) variation of divergence of first zero point with the order of scattered optical vortices. (b) The accuracy level of finding the topological charge using the divergence of the first zero point.}
    \label{fig:divergence}
\end{figure}

We have considered the divergence of the first zero point as a parameter to diagnose the order as one can easily observe the first ring for all spatial modes of light. We have studied the divergence of the first ring for different orders from $\ell=$0-8 and shown in Fig. \ref{fig:divergence}. One can clearly observe the divergence ($\alpha$) decays exponentially with the order of the optical vortex. This may be due to the exponential decay of the lowest root of Laguerre polynomial $c_\ell$. We found the divergence of the first ring experimentally and compared it with the theoretical results obtained by Eq. \ref{eq:div}. The figure shows that the experimental results are in excellent agreement with the theoretically obtained results. We also defined the level of accuracy for the measurement of topological charge of spatial modes using experimental ($\alpha_{ex}$) and theoretical ($\alpha_{th}$) divergence values as ($\alpha_{ex}-\alpha_{th})/\alpha_{ex}$. We found that one will be able to detect the spatial mode information of the perturbed vortices with the accuracy of more than 99 \% as shown in the right of Fig \ref{fig:divergence}.

In conclusion, we have provided an alignment-free technique for diagnosing the order of a perturbed vortex using the number of dark rings present in the 2-D spatial autocorrelation function. For higher-order spatial modes, we have utilized the divergence of the first ring for diagnosing the order as the observation of the number of dark rings is difficult. We have also provided the exact analytical expressions for both the autocorrelation function and divergence of the first ring and compared them with our experimentally obtained results. The results have a perfect match with each other. We will be able to get complete information about the mode with a small portion of the speckle pattern. The obtained results may find applications in free-space optical communication or satellite communication, where the diagnosis of spatial mode is essential.

SGR acknowledges the fund from SERB-DST through start-up research grant SRG/2019/000957. 

The authors declare no conflicts of interest related to this article.

\bibliography{References}

\end{document}